\documentclass[cameraready]{Interspeech}
\title{PF-D2M: A Pose-free Diffusion Model for Universal Dance-to-Music Generation}

\author[affiliation={1}]{Jaekwon}{Im}
\author[affiliation={2}]{Natalia}{Polouliakh}
\author[affiliation={2}]{Taketo}{Akama}
\address{
    $^1$ Graduate School of Culture Technology, KAIST, South Korea \\
    $^2$ Sony Computer Science Laboratories, Tokyo, Japan
}

\email{jakeoneijk@kaist.ac.kr, nata@csl.sony.co.jp, taketo.akama@gmail.com}


\usepackage{comment, makecell, multirow, amssymb}

\begin{document}

\maketitle

\begin{abstract}
Dance-to-music generation aims to generate music that is aligned with dance movements. Existing approaches typically rely on body motion features extracted from a single human dancer and limited dance-to-music datasets, which restrict their performance and applicability to real-world scenarios involving multiple dancers and non-human dancers. In this paper, we propose PF-D2M, a universal diffusion-based dance-to-music generation model that incorporates visual features extracted from dance videos. PF-D2M is trained with a progressive training strategy that effectively addresses data scarcity and generalization challenges. Both objective and subjective evaluations show that PF-D2M achieves state-of-the-art performance in dance-music alignment and music quality. Samples of PF-D2M are available at \url{https://jakeoneijk.github.io/pfd2m_project}
\end{abstract}

\section{Introduction}
%introduce the task of d2m and recent advance
Dance-to-music generation aims to synthesize music synchronized with corresponding dance movements. This technique can be applied to support creative workflows for choreographers, performers, and digital content creators, especially with the rise of short-form video content. Unlike traditional approaches \cite{traditionalmethod} that align a given music and dance motion using time scaling and time warping, deep learning-based methods can generate original music conditioned on dance movements.

To extract temporal information, previous dance-to-music generation methods primarily utilize motion features extracted from a single human dancer, such as the 3D Skinned Multi-Person Linear model (SMPL) \cite{smpl}  based features or 2D body keypoints. Several approaches \cite{dancetomidi1, dancetomidi2, dancetomidi3} focus on generating symbolic representations, such as MIDI, conditioned on dance sequences. However, symbolic music representations can be limited in expressing the expressiveness and dynamics of real dance music. To address this limitation, other approaches \cite{d2mgan} aim to generate audio representations with fine-grained acoustic information. Recently, diffusion models \cite{ddpm} have been widely used to produce high-quality music \cite{loris, textinversion}.

Despite recent advances in dance-to-music generation, two important challenges remain. First, motion features extracted from a single human dancer are often insufficient to represent the temporal information in dance videos. In particular, in scenarios involving multiple dancers, choreographies are aligned with music based on the movements of multiple performers rather than a single performer. Even in cases with a single dancer, most previous methods focus on extracting coarse rhythmic features \cite{loris, textinversion}. Furthermore, pose estimation modules may result in jittering pose sequences and exhibit limited performance on non-human dancers, such as 2D animated characters. Second, public high-quality dance-to-music datasets are scarce. AIST++ \cite{aist++}, which is used as the primary training dataset in most previous works, provides only 60 unique songs, with dance videos exhibiting relatively simple backgrounds. This scarcity of datasets acts as a major bottleneck in building dance-to-music models capable of generating diverse and high-quality music. However, expanding training data through crawling remains challenging due to copyright issues. In addition, in-the-wild dance videos often have poor audio quality caused by camera-recorded sound or contain scenes unrelated to dance.

To overcome these limitations, we propose PF-D2M, a diffusion model for universal dance-to-music generation. PF-D2M can effectively generate diverse and high-quality music for general dance videos, extending beyond scenarios involving a single human dancer. Instead of relying on motion features extracted from a single human dancer, PF-D2M incorporates visual features extracted from the video using Synchformer \cite{synchformer}. Additionally, to address data scarcity, we introduce a training recipe that progressively trains the model to extend its capabilities. Our contributions are summarized as follows:

\begin{itemize}
  \item We propose PF-D2M, a universal dance-to-music generation model capable of generating music for general dance videos, including scenarios with multiple dancers and non-human dancers.
  \item We introduce a progressive training strategy that effectively mitigates overfitting caused by data scarcity and improves the model’s generative capability.
  \item Both objective and subjective evaluations demonstrate that PF-D2M achieves state-of-the-art performance compared to previous methods.
\end{itemize}

\section{Method}
\begin{figure}[t]
  \centering
  \includegraphics[width=\linewidth]{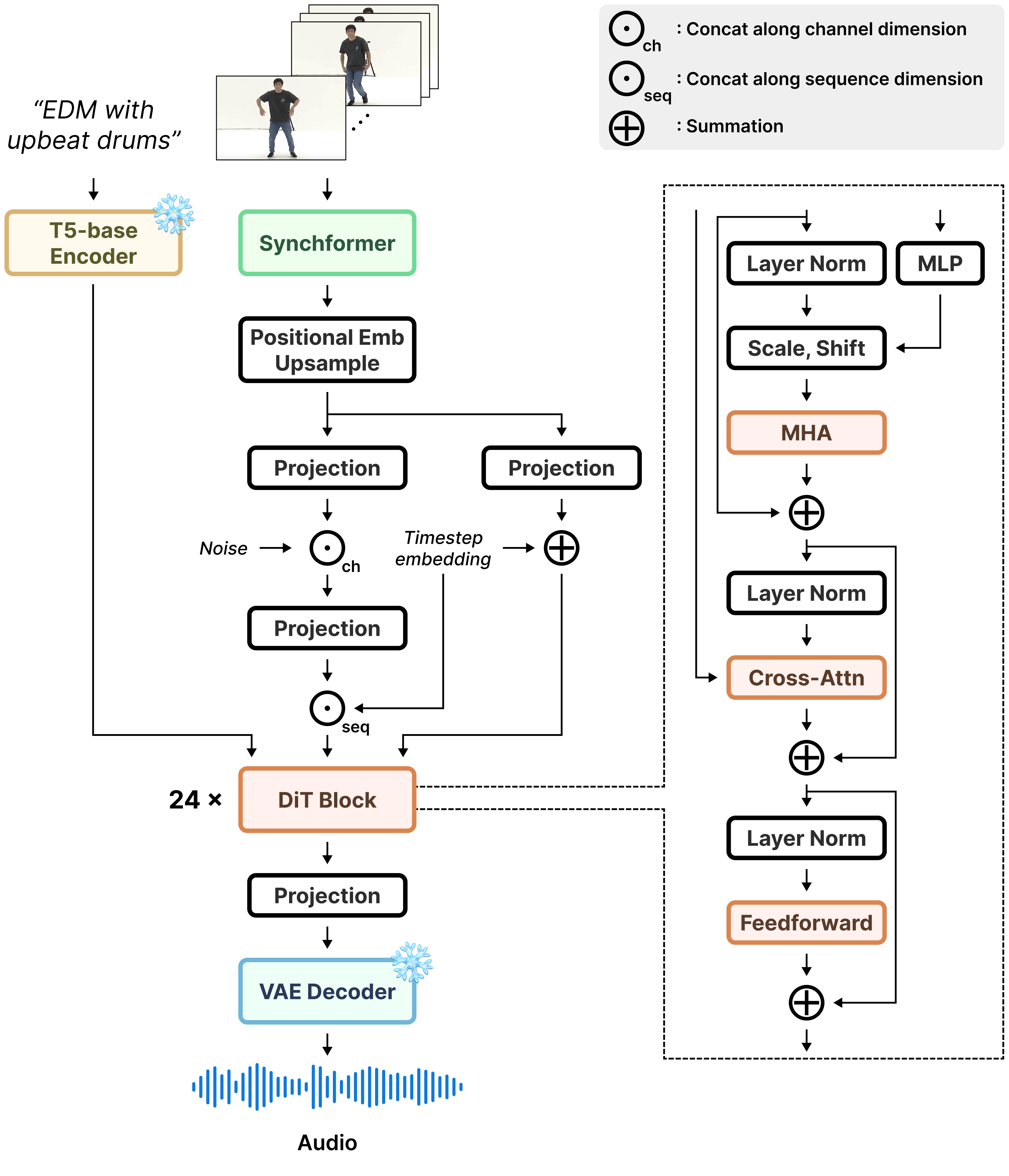}
  \caption{Overview of PF-D2M.}
  \label{fig:model_architecture}
\end{figure}
\subsection{Overview}
Figure \ref{fig:model_architecture} shows the overall architecture of PF-D2M. Let $\mathbf{v} \in \mathbb{R}^{(T \times f_v)\times H \times W \times 3}$ denote a silent dance video, where $T$, $f_v$, $H$, and $W$ are the duration, video frame rate (FPS), height, and width, respectively, and $3$ corresponds to RGB channels. Audio $\mathbf{a} \in \mathbb{R}^{2 \times (T \times f_s)}$ is the stereo music temporally aligned with $\mathbf{v}$, where $f_s$ is the audio sampling rate. The text caption $\mathbf{c}$ describes the music, such as genre, instruments and mood. PF-D2M generates $\mathbf{a}$, given $\mathbf{v}$ and $\mathbf{c}$ as input conditions. In Section \ref{method:dit_model}, we present the architecture of PF-D2M. In Section \ref{method:training_recipe}, we describe the training recipe composed of two stages.

\subsection{Model Architecture}
\label{method:dit_model}
We adopt the DiT architecture and the pre-trained VAE from \cite{stableaudioopen}. The VAE encoder compresses $\mathbf{a}$ into a latent representation $\mathbf{z} \in \mathbb{R}^{64 \times (T \times f_s)/2048}$. The DiT generates $\mathbf{z}$ given three conditioning features. First, text embeddings extracted from $\mathbf{c}$ using a pretrained T5-base encoder \cite{t5} are provided to the DiT through cross-attention. Second, sinusoidal embeddings of the diffusion timestep \cite{ddpm} are prepended to the input of the DiT. Third, visual features are extracted from $\mathbf{v}$ using the visual encoder of Synchformer \cite{synchformer} and added to learnable positional embeddings, following \cite{mmaudio}. To match the time dimension of $\mathbf{z}$, visual features are upsampled by nearest neighbor interpolation. These upsampled visual features are projected by a 1D convolution layer to match the channel dimension of $\mathbf{z}$ and then concatenated with the input of the DiT along the channel axis. Because we found that providing visual features solely through concatenation leads to slow convergence, an additional conditioning method is employed. The upsampled visual features are projected by a linear layer to match the hidden size of the DiT and are then summed with the timestep sinusoidal embeddings. These features are conditioned on every layer of the DiT through frame-wise scales and biases in adaptive layer normalization (AdaLN) \cite{adaLN} layers, similar to \cite{mmaudio}. Note that the first frames are excluded from AdaLN modulation, as they are treated as prepended global features.

The DiT is trained with the velocity-prediction objective \cite{vpred}. Both visual features and text embeddings are dropped out with a rate of 10\%, enabling classifier-free guidance \cite{cfg}. At inference time, we adopt DPM-Solver++ \cite{dpmsolver} with 100 inference steps and a classifier-free guidance scale of 5.0.

\subsection{Progressive Training Strategy}
\label{method:training_recipe}
We found two problems when training PF-D2M on the AIST++ dataset \cite{aist++}. First, the model is severely overfitted to the music in the training dataset. Due to the limited amount of music data provided by AIST++, the model tends to memorize the training music and lacks the ability to generate new music. Second, because AIST++ dance videos exhibit limited camera angles and background diversity, the model does not generalize well to real-world scenarios that often involve complex backgrounds, lighting conditions, and diverse viewpoints. To address these problems, we introduce a progressive training strategy composed of two stages.

\subsubsection{Stage 0: Text-to-Audio Pretraining Initialization}
Training a model capable of generating high-quality, diverse music requires substantial computational resources. In practice, we initialize the weights of PF-D2M with the weights of Stable Audio Open \cite{stableaudioopen}. Stable Audio Open is a text-to-audio model that demonstrates strong generative capabilities in both music and sound effects. To preserve its audio generation capabilities, the weights of modules only used in PF-D2M are initialized to zero. With this initialization strategy, PF-D2M performs identically to Stable Audio Open, except that it does not use timing conditions and operates on a different audio sequence length.
\subsubsection{Stage 1: Video-to-Audio Alignment Training}
Before training the model with music data, we first train it on VGGSound \cite{vggsound} to learn audio-visual synchronization. VGGSound provides approximately 500 hours of video data with diverse categories. This diversity enables the model to learn audio-visual synchronization in in-the-wild scenarios. For text captions, we use captions generated by Qwen-Audio \cite{qwenaudio1}, provided by AudioSetCaps \cite{audiosetcaps}.

Among the 309 audio classes provided by VGGSound, only a few classes, such as \textit{tap dancing} and \textit{people shuffling}, are related to dance-to-music. Their proportion is small (less than 10 hours), and many samples in these classes exhibit issues, including poor music quality due to field recordings, music consisting only of percussive sounds, and mislabeled videos. In addition, the diversity of both movement patterns and music genres is highly limited. 

Nevertheless, we observe that the model trained after Stage 1 is able to generate music across diverse genres for a wide range of dance videos, benefiting from text-to-audio pretraining initialization and video-to-audio alignment training on diverse video data. However, this model exhibits two major limitations. First, instead of generating musically coherent outputs, it tends to produce abrupt changes in the music that closely follow visual variations, resulting in a sequence of short musical excerpts rather than a single, consistent music track. Second, the generated audio often reflects room or environmental characteristics present in the video. Since our goal is to generate studio-quality music independent of the dancer’s location, this tendency does not align with our objective.

\subsubsection{Stage 2: Dance-to-Music Fine-Tuning}
To address the limitations of the Stage 1 model, we fine-tune it on AIST++ \cite{aist++}, a dance-to-music dataset provided as part of the LORIS benchmark \cite{loris}. Inspired by previous works \cite{mmaudio, multifoley} in video-to-audio tasks, we train the model on multimodal datasets to improve music generation capability and mitigate overfitting on AIST++. In Stage 2, the model is trained on the dance-to-music dataset (AIST++), text-to-music datasets (FMA \cite{fma}, MoisesDB \cite{moisedb}), and the video-to-audio dataset (VGGSound). Because each dataset has a different total audio duration, and to make the model focus more on music generation rather than sound effects, we sample training batches using a proportion of 2:4:1 for the dance-to-music, text-to-music, and video-to-audio datasets, respectively. For the text-to-music datasets, learnable empty embeddings are used for the visual features.

To generate text captions for the dance-to-music and text-to-music datasets, we employ Qwen2-Audio \cite{qwenaudio2}. We use the music tagging prompt from \cite{yue} to annotate genre, instrument, and mood tags. When Qwen2-Audio fails to generate appropriate tags, we utilize the metadata provided with each dataset. During training, we generate text prompts from annotated tags. Specifically, we construct captions by stochastically combining tags with predefined textual templates. When there is no valid tag available, we use a generic prompt such as \textit{``An instrumental music track"}. This randomized caption generation improves generalization, allowing the model to generate high-quality music regardless of the level of detail in the text caption.

\section{Experiments}
\subsection{Datasets}
\subsubsection{Text-to-Music Dataset Filtering}
PF-D2M focuses on generating danceable music without vocal content. For this purpose, we use approximately 14 hours of music from MoisesDB, created by summing instrumental stems without vocals. For FMA, we first exclude music belonging to four genres that are difficult to dance to or are likely to contain singing voice: Experimental, Folk, Old-Time, and Spoken. To further remove low-bandwidth audio, we compute the normalized spectral roll-off frequency \cite{sagasr} for each track and discard samples with values below 0.6. We then filter out music containing singing voice. Specifically, we first separate vocal components using the music source separation model \textit{htdemucs\_ft} \cite{demucs}, and subsequently apply Silero \cite{silero}, a voice activity detector, to determine the presence of singing voice. After filtering, the resulting dataset contains approximately 191 hours of audio.
\subsubsection{Preprocessing}
The video and audio are resampled to 25 fps and 44.1 kHz, respectively. All video and audio were randomly segmented into 7.98-second clips for training. Each RGB frame of the video is resized and center-cropped to $224\times224$.

\subsection{Implementation Details}
For Stage 1, PF-D2M is trained for 200{,}000 steps using a base learning rate of $1\times10^{-5}$ with a linear warm-up schedule for the first 1{,}000 steps. After 30{,}000 training steps, the learning rate is reduced to $1\times10^{-6}$. For Stage~2, PF-D2M is trained for 1{,}500 steps with a fixed learning rate of $1\times10^{-6}$. We use a batch size of 128 and the AdamW optimizer with $\beta_1 = 0.9$ and $\beta_2 = 0.999$ for all stages.

\subsection{Evaluation}
\subsubsection{Objective Evaluation}
We compare PF-D2M with state-of-the-art models, CDCD \cite{cdcd}, LORIS \cite{loris}, and Text-Inv \cite{textinversion}. For all comparison models, the official implementations are used. 

We evaluate our method and baseline models on AIST++ \cite{aist++} under the LORIS benchmark \cite{loris}. Following previous work, we reserve 5\% of the dataset for testing. However, unlike prior approaches that use random splits, we explicitly select specific music tracks for the test set: \textit{mBR0}, \textit{mMH0}, \textit{mLO2}, and \textit{mJB5}. This is motivated by two considerations. First, it ensures that the test set contains only music tracks that are unseen during training. Second, the test set should cover diverse music genres and tempos.

Following previous work, we use five objective metrics \cite{loris} to evaluate rhythm correspondence between the generated music and the ground-truth music. Specifically, we adopt Beat Coverage Score (BCS), Coverage Standard Deviation (CSD), Beat Hit Score (BHS), Hit Standard Deviation (HSD), and F1 score.

\subsubsection{Subjective Evaluation}
To evaluate in-the-wild scenarios, we collected challenging dance videos with diverse numbers of dancers, dance styles, dancer types, backgrounds, lighting conditions, and camera angles. We define four categories and collect five samples for each category, resulting in a total of 20 samples: \textit{single human dancer}, \textit{single non-human dancer}, \textit{multiple human dancers}, and \textit{multiple non-human dancers}. We define non-human dancers as performers that are not actual humans, including both 2D and 3D characters. %We define non-human dancers as dancers that are not actual humans. This category includes both 2D and 3D characters, whether or not they resemble human appearance. 
Both \textit{single human dancer} and \textit{single non-human dancer} categories refer to videos in which only one dancer performs, while additional people may appear as an audience. In contrast, \textit{multiple human dancers} and \textit{multiple non-human dancers} categories refer to videos in which multiple dancers perform, with the number of dancers ranging from 2 to 91.

For subjective evaluation, we compare PF-D2M against LORIS \cite{loris} and Text-Inv \cite{textinversion}. To obtain 2D skeletons for both LORIS and Text-Inv, we employ HRNet \cite{hrnet} from MMPose \cite{mmpose}, which is the same pose estimator used in both works. For fair comparison, all generated audio is segmented into 5.12-second clips, corresponding to the generation duration of Text-Inv, which has the shortest generation duration among the compared methods.

We conducted a subjective evaluation through a listening test with 20 participants. Participants were asked to assess dance-music alignment and music quality. Dance-music alignment measures how well the generated music matches the dance, regardless of music quality. Music quality evaluates the perceptual quality of the generated music, including audio fidelity and musical composition, when considered independently of the accompanying dance. For each question, participants evaluated the generated music from compared methods, rating each criterion on a scale of 1 to 5 points.

\section{Results}
\subsection{Objective Evaluation}
\begin{table}[t]
\caption{Objective evaluation results on the AIST++ test set. S1 denotes Stage 1 training, and S2 denotes Stage 2 training. %Higher BCS, BHS, and F1 and lower CSD and HSD correspond to better performance. 
The best result for each metric is shown in bold.}
\label{table:objective}
\centering
\small{
\begin{tabular}{c|c|c|c|c|c}
\Xhline{3\arrayrulewidth}

% ---- Metric row ----
Method
 & BCS↑ & CSD↓ & BHS↑ & HSD↓ & F1↑\\
\Xhline{3\arrayrulewidth}

% ---- Data rows ----
%model name & 0.00 & 0.00 & 0.00  \\
CDCD \cite{cdcd}& 89.2 & 9.0 & 93.8 & 10.0 & 91.5 \\
LORIS \cite{loris}& 89.9 & 8.9 & 95.3 & 8.9 & 92.5 \\
Textual-Inv \cite{textinversion}& \bf{90.6} & 11.1 & 80.9 & 28.3 & 85.5\\
\hline
PF-D2M (S1) & 90.5 & 13.1 & 91.2 & 18.6 & 90.9 \\
PF-D2M (S2) & 89.4 & \bf{8.1} & \bf{99.8} & \bf{1.9} & \bf{94.3} \\
\Xhline{3\arrayrulewidth}

\end{tabular}
}
\end{table}

\begin{figure*}[t]
  \centering
  \includegraphics[width=\textwidth]{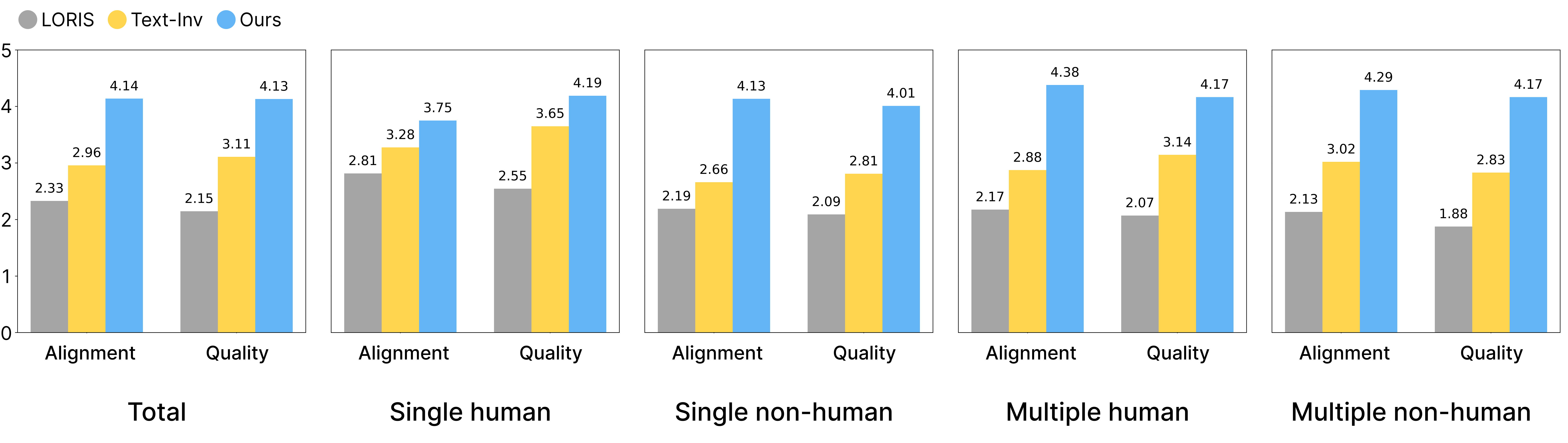}
  \caption{Subjective evaluation results on the in-the-wild set.}
  \label{fig:subjective_results}
\end{figure*}

Table \ref{table:objective} shows the results of the objective evaluation. PF-D2M achieves state-of-the-art performance on all metrics except BCS. These results demonstrate the effectiveness of incorporating Synchformer visual features, which yield better dance-music synchronization than methods relying on coarse rhythmic features from a single human dancer.

The Stage 2 model surpasses the Stage 1 model across all metrics except BCS. While the Stage 1 model lacks musical structural consistency, the Stage 2 model generates music with a coherent musical structure and still reflects fine-grained dance movements. %These results indicate the effectiveness of the dance-to-music fine-tuning stage in yielding generated music with more stable rhythm. %These results indicate that although the Stage 1 model can generate music aligned with dance videos, the dance-to-music fine-tuning stage is essential for improved dance-music synchronization. 
Although music quality is not captured by rhythmic evaluation metrics, it is worth noting that the Stage 2 model provides a better listening experience by generating high-quality music regardless of the dancer’s location, whereas the Stage 1 model often produces music that sounds like field-recorded audio.

PF-D2M achieves slightly worse performance on BCS compared to Text-Inv and LORIS. We observe that PF-D2M often generates music with subdivided rhythmic structures, such as dense hi-hat patterns, particularly in break and hip-hop genres. Although this rhythmic structure is appropriate for these genres, the increased number of detected beats can negatively affect the BCS metric. 

We observe that the objective metrics BCS, CSD, BHS, HSD, and F1 do not always correlate well with perceptual quality. Although there is no single correct music track for a given dance, these metrics evaluate generated music by comparing it to ground-truth music. As a result, outputs with rhythmic structures that differ from the ground truth may receive low scores, even when they are well aligned with the dance. Moreover, these metrics are not designed to assess audio fidelity or musical structure. Although some previous works \cite{fad} adopt Frechet Audio Distance (FAD) \cite{fad} to evaluate the quality of generated music, we do not use it because the test set is too small to yield reliable FAD results. Therefore, we put greater emphasis on subjective evaluation, leaving the development of more suitable objective datasets and evaluation metrics for dance-to-music generation as future work.

\subsection{Subjective Evaluation}
Figure \ref{fig:subjective_results} presents the results of the subjective evaluation. PF-D2M significantly outperforms the comparison methods across all metrics and test cases, demonstrating its effectiveness in in-the-wild scenarios. We observe that music generated by LORIS frequently lacks rich musical content, resulting in monotonous instrumental compositions and arrangements, while Text-Inv often produces audio with degraded sound quality. Moreover, both LORIS and Text-Inv often fail to capture fine-grained dancer movements, exhibiting weak alignment between the dance and the generated music. In contrast, PF-D2M generates high-quality music that reflects detailed dance movements. Notably, PF-D2M often produces music arrangements that highlight specific movements, making the dance content more dynamic and expressive.

PF-D2M further increases the performance gap compared to baseline methods in scenarios beyond the single human dancer setting, such as non-human dancer and multiple-dancers cases. Furthermore, we found that PF-D2M demonstrates robustness under complex visual conditions, including non-typical camera viewpoints and multi-cut videos, whereas the comparison methods often fail to achieve proper dance-music alignment. These results highlight the versatility of PF-D2M and its stronger applicability to real-world scenarios.

\section{Conclusion}
In this paper, we propose PF-D2M, a diffusion model for universal dance-to-music generation. By incorporating visual features extracted from dance videos, PF-D2M avoids reliance on single human pose representations and supports a broader range of scenarios, including multiple dancers and non-human dancers. We further introduce a progressive training strategy that combines text-to-audio initialization, video-to-audio alignment training, and dance-to-music fine-tuning, which effectively mitigates overfitting issues caused by data scarcity. Both objective and subjective evaluations demonstrate that PF-D2M achieves state-of-the-art performance in dance-music alignment and perceptual music quality.

Although PF-D2M demonstrates strong performance across various scenarios, several limitations remain that warrant future work. First, the duration of the music generated by PF-D2M is relatively short compared to real-world use cases. Future work could explore generating longer-duration music that is well aligned with diverse dance progressions. Second, the lack of an evaluation dataset for dance-to-music generation hinders objective evaluation, highlighting the need for benchmarks with diverse music and dance videos.

\bibliographystyle{IEEEtran}
\bibliography{mybib}

\end{document}